# A Possible Method of Carbon Deposit Mapping on Plasma Facing Components Using Infrared Thermography

**R. Mitteau**, J. Spruytte, S. Vallet, J.M. Travère, D. Guilhem, C. Brosset

*Association Euratom - CEA sur la Fusion Contrôlée, direction des sciences de la matière, Centre d'Etude de Cadarache F-13108 Saint Paul Lez Durances CEDEX*

## Abstract

The material eroded from the surface of plasma facing components is redeposited partly close to high heat flux areas. At these locations, the deposit is heated by the plasma and the deposition pattern evolves depending on the operation parameters. The mapping of the deposit is still a matter of intense scientific activity, especially during the course of experimental campaigns. A method based on the comparison of surface temperature maps, obtained in situ by infrared cameras and by theoretical modelling is proposed. The difference between the two is attributed to the thermal resistance added by deposited material, and expressed as a deposit thickness. The method benefits of elaborated imaging techniques such as possibility theory and fuzzy logics. The results are consistent with deposit maps obtained by visual inspection during shutdowns.

## 1. Introduction

Tokamaks with carbon as main plasma facing material are increasingly encountering issues with deposits in areas which are close to the main plasma surface interaction zones [refs. 1-3]. Concerns include topics related to deuterium retention and the interpretation of the infrared images of hot surfaces on divertor and limiters. The measure of the deposit thickness is complicated by constraints of accessibility, availability of the data and required precision. Quarks microbalances are not usable in these areas facing the plasma, and optical methods (interferometry [4], radar [5]) appear as the most appropriate. However, they are still under development, and a method based on an existing diagnostic delivering available data is looked after.

Deposits are known to influence the surface temperature of plasma facing components during plasma discharges, by adding a thermal resistance at the surface. As a consequence, surfaces with a thick deposit appear hot, and even hotter than those subject to the peak heat flux. In Tore Supra, a section of the toroidal pumped limiter (60° toroidally, composed of three sectors of 20°) is observed with an endoscope working in the infrared range. It is equipped with a focal plane array camera [6], and the films of the discharges are stored and available for analysis. Carbon deposit is observed and the issue of its characterisation is becoming more acute because of the interaction of the high temperatures observed with Tore Supra safety regulations. In 2005, the deposit temperature reaches up to 2000K with an additional power as low as 4 MW, whereas in 2004 up to 11.5 MW could be injected with deposit temperature of the order of 1200K. The deposit evolution is a probable consequence of Tore Supra unique characteristics of actively cooled wall, long pulses (custom 60 s - record 400 s) at high power (custom 4 MW - record 11.5 MW). Tore Supra explores basically steady state discharges, and deposit analysis is performed on the CW plateau, for which experimental data as well as plasma parameters do not depend on time.

The principle of the evaluation is to compare the limiter temperature map, measured by the infrared imaging system, to the one calculated using a heat flux model [7]. The temperature difference is attributed to the carbon deposit, and translated into a material thickness using an assumption on the deposit thermal conductivity. The main challenges of the method are imaging techniques, because the whole process has to be automatic.

## 2. Method

The method is based on an automated algorithm for the pattern recognition. It is chosen to work at the scale of the tile (25 x 25 mm²), because the endoscope diagnostic spatial resolution is of the order of 10 mm which is not sufficient to measure the tiles leading edges temperature (typical dimension of 1 mm). The algorithm uses the grid lines of the gaps between the tiles, apparent in large areas of the image. They are caused by hot carbon deposit on the leading edges of the tile gaps [8].

Following operations are applied to the image:

- Segmentation using a watershed algorithm. This operation discloses the valleys and the divides between catchment basins. This operation equalises the image levels, and selects simply the grid features.
- Deformation of the image, using a warping algorithm. This operation rectifies the grid, so that it is composed only of vertical and horizontal straight lines.
- Calculation of the grid intrinsic periods : this is done by summation of the columns and the lines, and by deducing the signal period.
- Determination of the grid origins. The operation is trivial in the poloidal direction, where the leading edge gives the starting point of the grid. Toroidally, a normalized correlation is used between a reference sample and the infrared image. The maximum gives the location of the origin.
- The complete grid can then be re-created using these results and poloidal and toroidal indexes created.
- The inverse deformation is then performed on the index images.

Two images of the same pixel count of the original one are obtained after these operations: the first image is the element number, the second one the tile number on the element (Fig. 1). A pixel value of zero in the index images means a limit between two tiles or a pixel outside the limiter.

The following step consists in calculating a best estimation of the tiles temperature. Due to the hot leading edges, a simple averaging of the pixel temperatures does not always suffice. In areas with hot leading edges, taking the local minimum temperature is a better estimation. Three tile groups are defined:

1. Tiles with low temperature
2. Tiles with high temperature caused by high heat flux
3. Tiles with high temperature caused by carbon deposit

The classification of the tiles relies on expert user's knowledge. The group affectation of each tile uses the possibility theory, developed for example in face recognition applications [9]. The possibility theory principle is to maintain all possibilities until the last decision step. The method is based on fuzzy logics. Five expert rules are defined, the 3 most important being (the <u>other two</u> are technical and are not listed):

a) Low temperature tiles are cooler than the limiter mean temperature.
b) The temperature of hot tiles caused by high heat flux is hotter on the edges than on the centre, and is not excessively hot.

c) High temperature tiles with hot borders of the same magnitude are probably tiles with deposit, even more if they are excessively hot.

The rules expressing the expert knowledge are based on attributes $R_1$ (normalised average temperature) and $R_2$ (convexity), which are tile dependant figures.

$$R^i_1 = \frac{\langle T^{center}_{pixels} \rangle_{tilei}}{\langle \langle T^{center}_{pixels} \rangle_{tilei} \rangle_{alltiles}} \quad , \quad R^i_2 = \frac{\langle T_{edge\,pixels} \rangle_{tilei}}{\langle T_{center\,pixels} \rangle_{tilei}}$$

Using the index images, the tile is described as a group of typically 10 pixels, surrounded by also typically 10 edge pixels. The attributes define the degree of belonging to the fussy sets. For example, a tile with $R_1 = 2.1$ and $R_2 = 1.3$ will belong to a certain extend to the fuzzy set b) (Fig. 2). Depending on the degree of belonging to the fuzzy set, the possibility is evaluated. In Fig. 2, the application of rule b is detailed for 3 tiles. It details how the possibility is upgraded to avoid a large influence on the classification when the rules are not sufficiently discriminating. The principle for the merging of the different rule possibilities is described in [9]. A quality indication of the sorting can be deduced from the difference between the highest possibility to the following one, an interesting feature of the method. A difference of 1 indicates a doubtless sorting.

As a result, a map of the best possible estimation of the tile temperature is obtained (relative to the expert knowledge). It is worth mentioning that this is a kind of optimum compression of the image, replacing an array of 76800 values by 2016, a reduction of the data amount by a factor of 38 (compression of 97%).

The limiter surface temperature modelling is done with the heat flux code TOKAFLU. It is based on a standard exponential scrape off layer model, and expresses the heat flux as a product between an exponential decaying parallel flux to the incidence angle. The model includes cross field heat flux, adjusted to match the temperature pattern for cases with negligible carbon deposit [10]. The supplementary limiter heat flux caused by ions trapped in the ripple wells during ion cyclotron resonance heating has also been included [11]. The resulting heat flux is simply translated into a surface temperature using the univalent relation linking the two for plasma facing elements working in steady state.

## 3. Results and discussion

The algorithm has been investigated and updated during the 2005 experimental campaign. The work focused on endoscope data of sector 3 (Q3) and the section investigated is PJ4. The first lesson is related to the expert rules use and adjustment, and the setting of the degree of possibilities. An important asset of the technique is that once the principles have been set, the algorithm can be easily improved with further rules without requiring more complex coding. The regular use of the sorting leads to propose the implementation of a new rule, stressing that there should not be isolated small group of tiles of one category (the tiles categories are connected within simple sets of only one or two per image.)

The resulting carbon deposit map is given in fig. 3 for the pulse 34663, during the plateau phase with a plasma current of 1.1 MA. The deposit appears in white shades. The white dashed contour indicates in a binary approach the area where deposit is evidenced (visual inspection during the 2005-2006 winter shut down). There is a good geometrical agreement between the map and the contour. The peak deposit areas are in the tips of the private flux region, which is coherent to previous investigations [10]. An area of thin deposit appears on the low field side (at the bottom of the limiter in fig 3), where the heat flux is low. There is few change of this carbon map with variable plasma parameters, indicating that the deposit is steady. This is coherent with the hour long time scale of the deposit formation [8], much longer than the one of an operational session.

The deposit conductivity is set here to 0.01 W/mK so that the peak thickness matches the in situ estimation of 200µm. This value is 2 orders of magnitude smaller than a previous estimation of 1 W/mK obtained on the neutraliser element with a 800µm thick deposit [12]. Two explanations can be advanced to explain this discrepancy: either the limiter deposit thermal conductivity is indeed much lower than the one measured on the neutraliser in 2003 (which could be due to the fragmentation and flaking of the deposit), or the large overheating observed is actually a consequence of an underestimating of the endoscope optics transmission. The calibration procedure for the endoscope transmission is currently being reviewed to assess this point.

The deposit thickness is plotted against the tile number for the 5$^{th}$ element of the PJ4 sector in Fig. 4, along with the data from IR interpretation and temperature modelling. The agreement is good between the calculated and measured temperatures in the heat loaded area. The highest temperatures are recorded just inside of the private flux area, where only radiation is expected. This encompasses 2 adjacent tiles, where the deposit thickness is 70 µm. This area, neighbour of the frontier between the heat loaded and private flux region, is a preferential area of material deposition. Under variable plasma conditions (mainly $q_{edge}$), the precise location of the frontier changes so that this deposition area can be wetted by the plasma, and heat up considerably. Another observation is related to the deposit thickness in the centre of the private flux region (tiles 2-7 on element 5), measured to 5µm, whereas is it null in the erosion area (tiles 10 – 19). This is coherent with our understanding of the erosion/redeposition mechanisms on the limiter, and the thickness of 5 µm is also coherent with a raw estimation of 10 µm, based of the fact that the woven structure of the carbon composite is still visible under the deposit.

It should be stressed that the quantitative results of the method are highly dependant on the quality of the data delivered by the infrared diagnostic, which is strongly dependant on the measure of the infrared optics transmission. The Tore Supra infrared imaging diagnostic is currently being improved technically to have a direct and independent measure of this transmission. The evaluated deposit thickness is also highly dependant on the assumption made on the thermal conductivity of the deposit, a parameter not easily measured and which may vary significantly depending on location and time. The carbon deposit thermal conductivity is investigated in several experiment, and is to be consistently cross checked.

## 4. Conclusion

The method has proven the capability to produce a carbon deposit map which is coherent with one obtained from an independent method. It is a demonstration of the capability of imaging techniques applied to an infrared diagnostic. The algorithm has been largely automated, however manual tuning is still mandatory for some aspects. Nevertheless, the method is very promising, and would be of great interest to be used as automatic data processing, to produce "meta-data" from the infrared diagnostic. Finally, an effort must be set on the infrared diagnostic calibration and deposit thermal characterisation, if the method is to be used fully quantitatively.

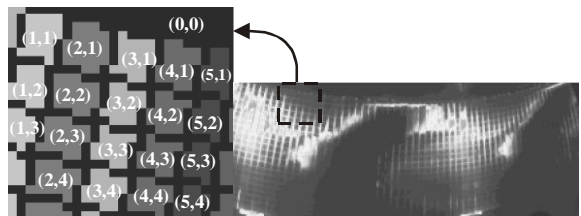

Fig. 1 : Indexation of the tile : the left figure is the element index, the right one the index of the tile on the element.

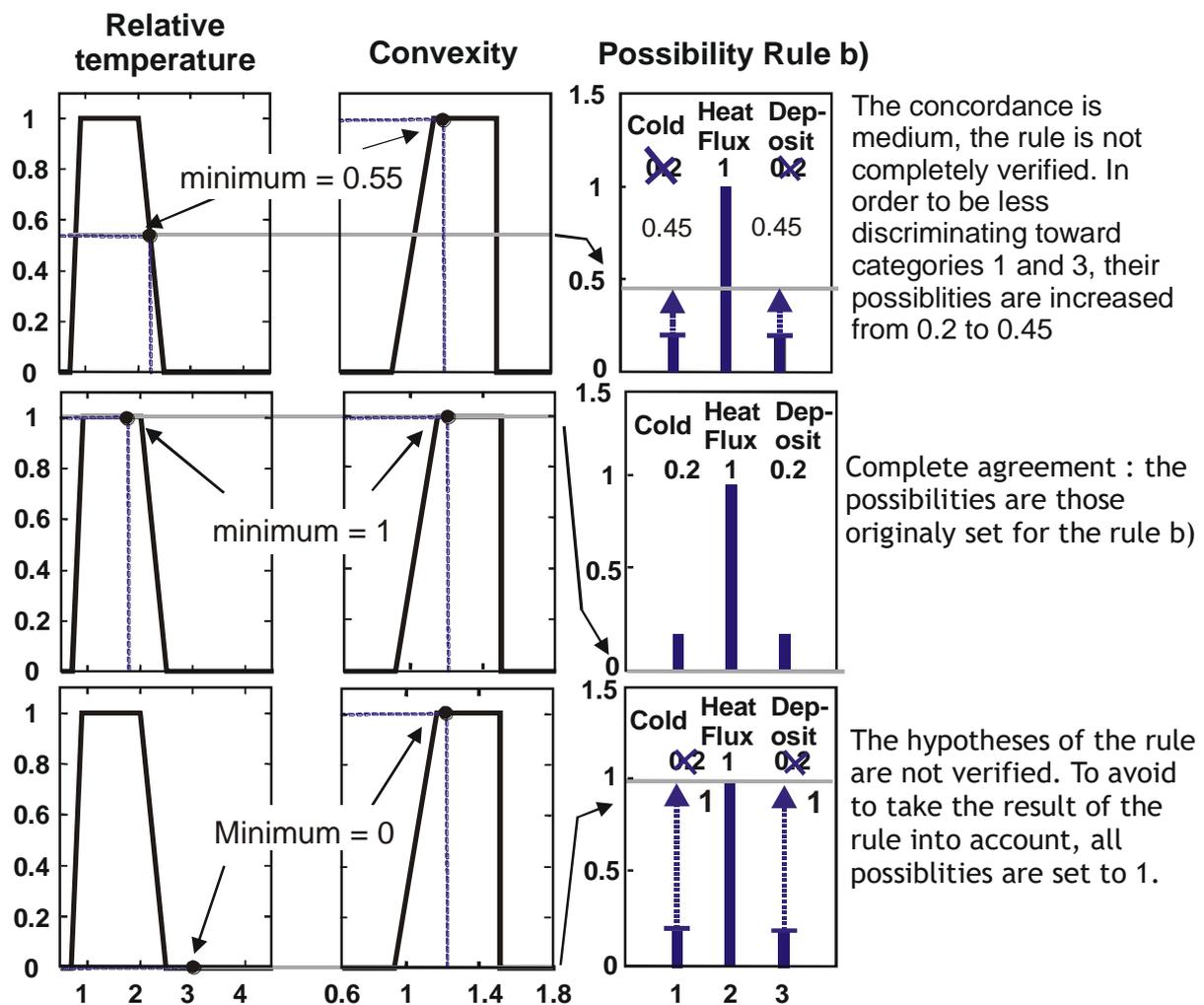

Fig. 2 : Example of the setting of the possibility for the expert rule b)

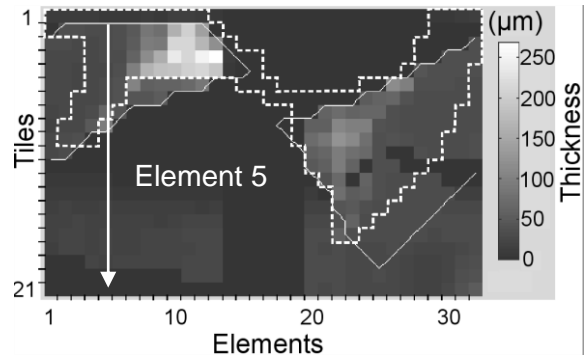

Fig. 3 : Map of the deposit on a section of 20° of the toroidal pump limiter. The colormap on the right hand side is the deposit thickness expressed in micrometers. The thin white contour is the limit of the private flux area

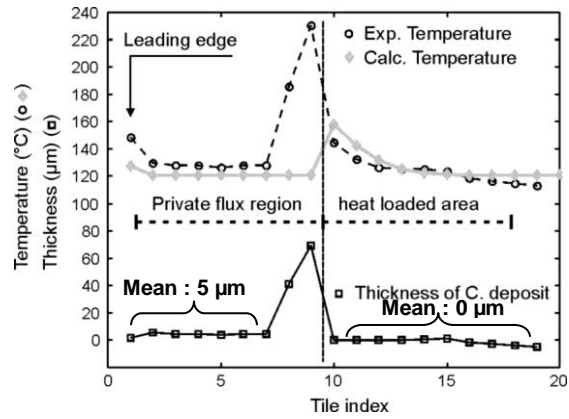

Fig. 4 : Temperature and deposit thickness along element 5

# POSSIBLE METHOD OF CARBON FACING COMPONENTS USING

R. Mitteau, J. Spruytte, S. Vallet, J.M. Travère, D. Guilhem, C. Brosset

## Overview : *From an infrared image to a carbon deposit thickness evaluation*

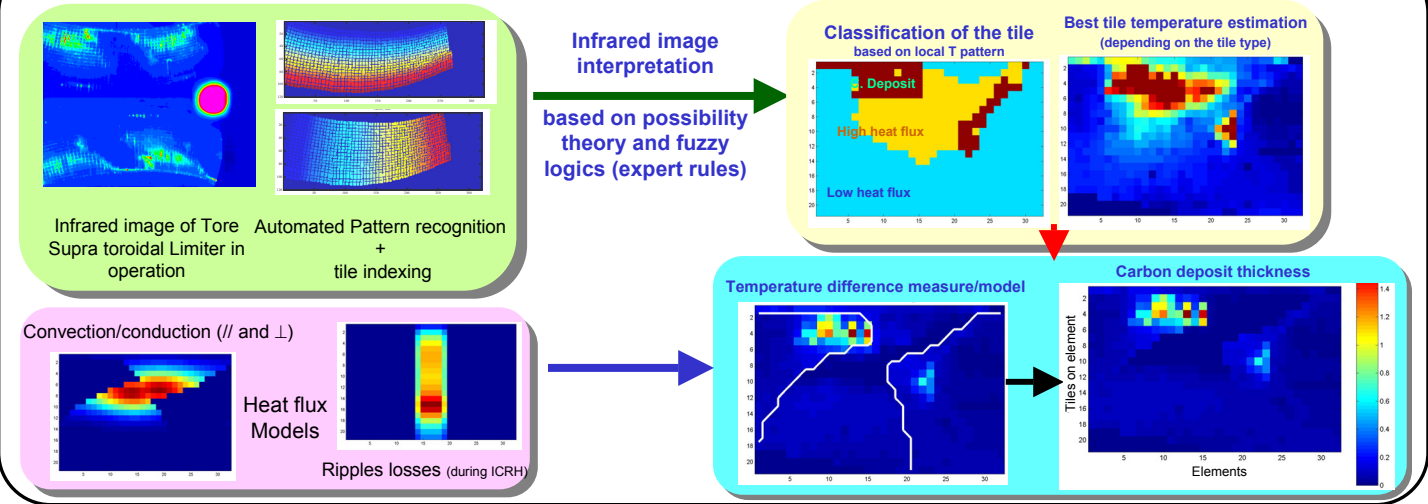

## Introduction : *carbon deposit observations and consequences on infrared measurements*

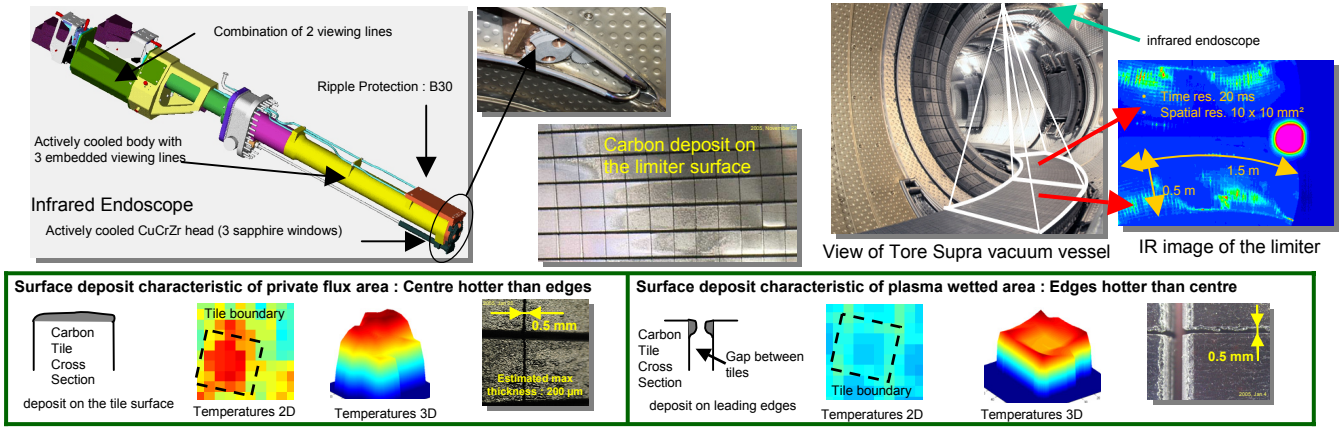

⇒ The best estimation for the tile temperature is not always the average value : the local minimum is a better estimation in high heat flux areas

## Method for tile indexation

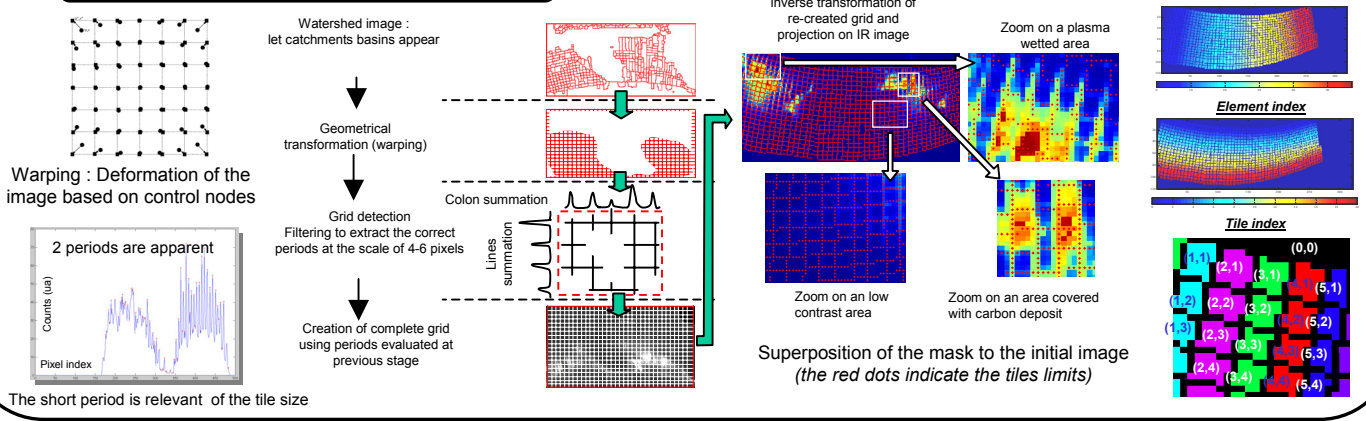

# DEPOSIT MAPPING ON PLASMA INFRARED THERMOGRAPHY

## Association EURATOM-CEA
CEA / DSM / Département de Recherches sur la Fusion Contrôlée
CEA-Cadarache, 13108 ST PAUL-LEZ-DURANCE (France)

## Method for tile classification and temperature evaluation

Classification by *possibility theory and fuzzy logics*.
Expert knowledge : classification rules (the first 3)
a) Low temperature tiles are cooler than the mean temperature of the limiter
b) The temperature of high temperature tiles caused by high heat flux is hotter on the edges than on the centre, and are not excessively hot.
c) High temperature tiles with hot borders of the same magnitude are probably tiles with deposit, even more if those temperatures are excessively hot.

Data :
- Tile pixels mean temperature
- edge pixels mean temperature
- mean tile temperature over all tiles

Attributes :
- relative temperature (R1)
- convexity (R2)

Classes :
- Low heat flux tiles
- Tiles that are hot because of carbon deposit presence
- Tiles that are hot because of high heat flux

data → Attributes

Mean edge pixel value

R2 = (Mean edge pixel value) / (Mean tile pixel value)

Higher than 1 → R1
Slightly higher than 1 → R2

Distribution of the attribute for the tile set

Possibility for rule b
Cold 0.2  Heat flux 1  Deposit 0.2

Determination of the possibility :
Setting the degree of concordance

Relative temperature | Convexity | Possibility Rule b
minimum = 0.55 — The concordance is medium, the rule is not completely verified. In order to be less discriminating toward categories 1 and 3, their possibilities are increased from 0.2 to 0.45.
Cold ½  Heat flux 1  Deposit ½   →  0.45 1 0.45

minimum = 1 — Complete agreement : the possibilities are those originaly set for the rule b)
0.2 1 0.2

Minimum = 0 — The hypotheses of the rule are not verified. To avoid to take the result of the rule into account, all possibilities are set to 1.
1 1 1

## Results

The analysis is made during steady state phase (>1s)

Power - MW vs Time - s

#34663, plateau at 1.1 MA — element 5 — Tiles vs Elements, Thickness (µm)

#34663, plateau at 1.1 MA
Leading edge | Private flux region | Heat loaded area
Exp. Temperature / Calc. Temperature
Thickness of C. deposit
Profiles along element 5 — Tile index

carbon deposit (hot)
carbon deposit (hot)
Hot areas caused by heat flux
Boundary of private flux area (from model)
Low temperature area

Element 16 17 32
Thin carbon deposit
Thick carbon deposit
Flakes
Heavy flaking
In situ observations January 2006 shut down sector Q3A
No data

- Mapping is coherent with in situ observation
- Carbon deposit thermal conductivity (0.01 W/mK) set so that actual max. deposit thickness matches in situ observation (2 order of magnitude smaller than previous estimate of 1 W/mK )
- Carbon deposit maps are steady with variable plasma

## Conclusions

- Carbon deposit maps are obtained
- Learning process : how to set the rules, indications toward further ones.
- Once the coding is done, adding further rules is simple.
- Demonstration of the capability of imaging techniques
- Interesting : the *degree of confidence* on the classification is given by the difference of the maximum possibility to the second one
- It is also an optimum data compression
- *But* deposits thickness much dependant on the IR optics calibration (measure of optics transmission – until this, only relative map)
- Still some work to facilitate the use of the algorithm (change of line of sight, auto check of the tile indexing…)